\documentclass[a4paper]{article}

\usepackage{INTERSPEECH2020}

\usepackage[utf8]{inputenc} 
\usepackage[T1]{fontenc}    
\usepackage{hyperref}       
\usepackage{url}            
\usepackage{booktabs}       
\usepackage{amsfonts}      
\usepackage{nicefrac}       
\usepackage{microtype}      
\usepackage{lipsum}
\usepackage{color}

\usepackage{hyperref}
\usepackage{url}
\usepackage{graphicx}
\usepackage{subcaption}
\usepackage{bm}
\usepackage{multirow,booktabs}
\usepackage{amsmath}

\title{Unconditional Audio Generation with Generative Adversarial Networks and Cycle Regularization}

\name{Jen-Yu Liu$^1$, Yu-Hua Chen$^{1,2}$, Yin-Cheng Yeh$^1$, Yi-Hsuan Yang$^{1,2}$}
\address{
  $^1$Taiwan AI Labs, Taipei, Taiwan\\
  $^2$Academia Sinica, Taipei, Taiwan}
\email{jyliu@ailabs.tw, cloud60138@citi.sinica.edu.tw, yyeh@ailabs.tw, yang@citi.sinica.edu.tw}

\begin{document}

\maketitle

\begin{abstract}
In a recent paper, we have presented a generative adversarial network (GAN)-based model for unconditional generation of the mel-spectrograms of singing voices. As the generator of the model is designed to take a variable-length sequence of noise vectors as input, it can generate mel-spectrograms of variable length. However, our previous listening test shows that the quality of the generated audio leaves room for improvement. The present paper extends and expands that previous work in the following aspects. First, we employ a hierarchical architecture in the generator to induce some structure in the temporal dimension. Second, we introduce a cycle regularization mechanism to the generator to avoid mode collapse. Third, we evaluate the performance of the new model not only for generating singing voices, but also for generating speech voices. Evaluation result shows that new model outperforms the prior one both objectively and subjectively.  We also employ the model to unconditionally generate sequences of piano and violin music and find the result promising. Audio examples, as well as the code for implementing our model, will be publicly available online upon paper publication. 
\end{abstract}

\section{Introduction}
\label{sec:intro}

In the recent development of deep learning, unconditional generation of images has been a popular research topic \cite{goodfellow14,pixelrnn,Karras2019}, either for its own artistic value or for being used as a base model for developing models with more fine-grained conditions (e.g., \cite{oord16nips}). Unconditional music generation in the symbolic domain, which aims at generating sequences of musical notes in a symbolic format such as the piano roll, has also become an active research topic lately \cite{musegan,huang2018music,huang20remitransformer}.\footnote{Following the convention in the literature, we define unconditional generation as a task that aims at generating things from scratch, i.e., taking nothing but random noises as the input. In contrast, a conditional generation model takes additional input such as class labels \cite{cgan}, text \cite{tacotron2}, pitch labels \cite{gansynth,wang2019performancenet}, or reference audio \cite{vc1,ddsp20iclr,jayneel20icassp}.}

Unconditional generation of raw audio waveforms, or its time-frequency representations, has received growing attention in recent years as well. 
A notable example is the \emph{Zero Resource Speech Challenge} (``TTS without T,'' or text-to-speech without text) organized in some editions of the INTERSPEECH conference \cite{ZR19}. Given a collection of raw audio without text or phoneme labels as the training data, the participants have to discover the subword units of speech \cite{chen2019completely,chorowskiWBO19,eloffNNGNPBWSK19} in an unsupervised way so as to synthesize novel utterances from novel speakers.
While being an interesting and meaningful unconditional audio generation task, the task setup is fairly speech-specific and accordingly such TTS without T models are not readily applicable to generate other types of audio signals, such as instrumental or environmental sounds.

There have been attempts to unconditional  generation of general audio, e.g., \cite{Dieleman2018,melgan,melnet}. However, they all use an auto-regressive approach that takes only a noise vector as input and generates sequentially samples of an audio signal (e.g., timesteps or spectral frames), one sample at a time, which might not be efficient at inference time. We are motivated to explore alternative model architectures.

We present such an attempt in a recent work for unconditional singing voice generation \cite{liu2019score}, aiming to generate improvised singing voices without using not only the phoneme labels (i.e., the lyrics) but also the pitch labels (i.e., the singing melody), in both model training and inference time.
It is based on a generative adversarial network (GAN) \cite{goodfellow14} where the generator takes a variable-length sequence of noise vectors as input, instead of just one noise vector as done in \cite{Dieleman2018,melgan,melnet}.
The mission of the generator is to convert the sequence of input noise vectors to a mel-spectrogram of the corresponding length, with each input vector corresponds to a certain length at the output.
Such an architecture therefore has the potential to generate a variable-length audio in a more efficient way.
While being an interesting attempt, the user study reported in the previous work (i.e., \cite{liu2019score}) suggests that the model presented there is not powerful enough to generated samples with satisfying perceptual quality.  Moreover, whether such a GAN-based model can be applied to audio other than singing voices remains unexplored. 

We presented in this paper an improved version of our prior model, using a similar GAN architecture that takes multiple noise vectors as input. But, the new generator now employs a hierarchical structure to govern the temporal coherence of the generated samples. Moreover, we regularize the model training process by enforcing cycle consistency between each input noise vector and the corresponding segment in the output mel-spectrogram. We validate both objectively and subjectively that the new model greatly outperforms the prior one, for unconditional generation of not only singing voices but also speech.

We refer to the proposed model architecture as UNAGAN, or \underline{un}conditional \underline{a}udio generation with GAN. The code and trained models are available at \url{https://github.com/ciaua/unagan.git}. Samples of the generated sounds of singing, speech, as well as instruments are also provided.

\section{Problem Formulation}
\label{sec:formulation}

In the literature, a common approach to audio generation is to first generate acoustic features, such as the mel-spectrograms, and then pass them to a vocoder to generate the corresponding audio waveforms \cite{tacotron2,melgan}. 
We follow this practice in this paper, and we focus on generating the mel-spectrograms.

The problem formulation adopted in \cite{Dieleman2018,melgan,melnet} can be generally described as follows. Given an input noise vector $\mathbf{z}\in \mathbb{R}^{N}$, where $N$ denotes the length of the vector, they build a generator $G(\cdot)$ so as to convert the input vector to a sequence of acoustic features (e.g., the mel-spectrograms),
$\widehat{\mathbf{X}} \equiv G(\mathbf{z}) \in \mathbb{R}^{K \times T}$, in an auto-regressive manner. Here, $K$  denotes the length of each acoustic feature vector $\widehat{\mathbf{x}}_t$, and $T$ the temporal length of the sequence to be generated. 

We intend to use instead a sequence of random noise vectors as the input, namely, $\mathbf{Z}=[\mathbf{z}_1, \mathbf{z}_2, \dots, \mathbf{z}_{T'}]$, where each term in each vector $\mathbf{z}_t$ is sampled from a Gaussian distribution with zero mean and unit variance. And, the length of the input sequence, namely $T'$, is proportional to the length of the target output sequence, namely $T$. In other words, each $\mathbf{z}_t$ has a direct influence over one (i.e., when $T'=T$) or a few (when $T'<T$) samples of the target output $\widehat{\mathbf{X}}$.

\section{Model}
\label{sec:model}

We adopt the GAN framework for unconditional generation of audio signals of arbitrary length. Similar  to \cite{liu2019score}, we find that the boundary-equilibrium GAN (BEGAN) \cite{berthelot17} works better than other types of GANs for this task, so we also use it here. To further improve the generation quality, we propose a number of changes for  the  generator and the discriminator, as described below.
We start with a brief introduction of BEGAN first.

\subsection{Boundary-Equilibrium GAN}
\label{sec:began}

In BEGAN \cite{berthelot17}, the loss functions $l_D$ and $l_G$ for the discriminator $D(\cdot)$ and the generator $G(\cdot)$ are respectively:
\begin{align}
    l_D=& L(\mathbf{X})-\tau_s L(G(\mathbf{Z})) \,, \label{eq:1}\\ 
    l_G=& L(G(\mathbf{Z})) \,, \label{eq:2} 
\end{align}
where $\mathbf{X} \in \mathbb{R}^{K \times T}$ denotes a sequence of acoustic features from a real audio signal sampled from the training data, 
and $L(\cdot)$ is a function that measures how well the discriminator reconstructs its input. Specifically,
\begin{equation}
L(\mathbf{M})=\frac{1}{WT}\sum_{w,t}|D(\mathbf{M})_{w, t}-\mathbf{M}_{w,t}| \,, 
\end{equation}
for an arbitrary $W\times T$ matrix $\mathbf{M}$,
where $M_{w,t}$ denotes the $(w, t)$-th element of a matrix $\mathbf{M}$ (and similarly for $D(\mathbf{M})_{w, t}$).
Moreover, the variable $\tau_s$ in Eq. (\ref{eq:1}) is introduced by BEGAN to balance the power of $D(\cdot)$ and $G(\cdot)$ during the learning process. It is dynamically set to be $\tau_{s+1}=\tau_s+\beta(\gamma L(\mathbf{X})-L(G(\mathbf{Z})))$, for each training step $s$, with $\tau_s \in [0,1]$. And, $\beta$ and $\gamma$ are manually-set hyperparameters. From Eqs.~(\ref{eq:1}) and (\ref{eq:2}), we see that $D(\cdot)$ and $G(\cdot)$ have contradicting goals, giving rise to the name of adversarial training. Once trained, only $G(\cdot)$ is used at the inference time for generating new content from scratch.

\subsection{Hierarchical Structure in the Generator}
\label{sec:generator}

In our previous work \cite{liu2019score}, each spectral frame $\widehat{\mathbf{x}}_t$ has its own input noise $\mathbf{z}_t$, i.e., $T'=T$. This allows the generator to generate a sequence of arbitrary length. However, this could also make it difficult for the generator to generate coherent sequences of acoustic features while respecting the input noises, because $[\mathbf{z}_1, \mathbf{z}_2, \dots, \mathbf{z}_{T'}]$ are sampled independently. 
As a result, the generated sounds tend to be fragmented and do not form complete pronunciations.

\begin{figure}
	\centering
	\includegraphics[width=0.8\columnwidth]{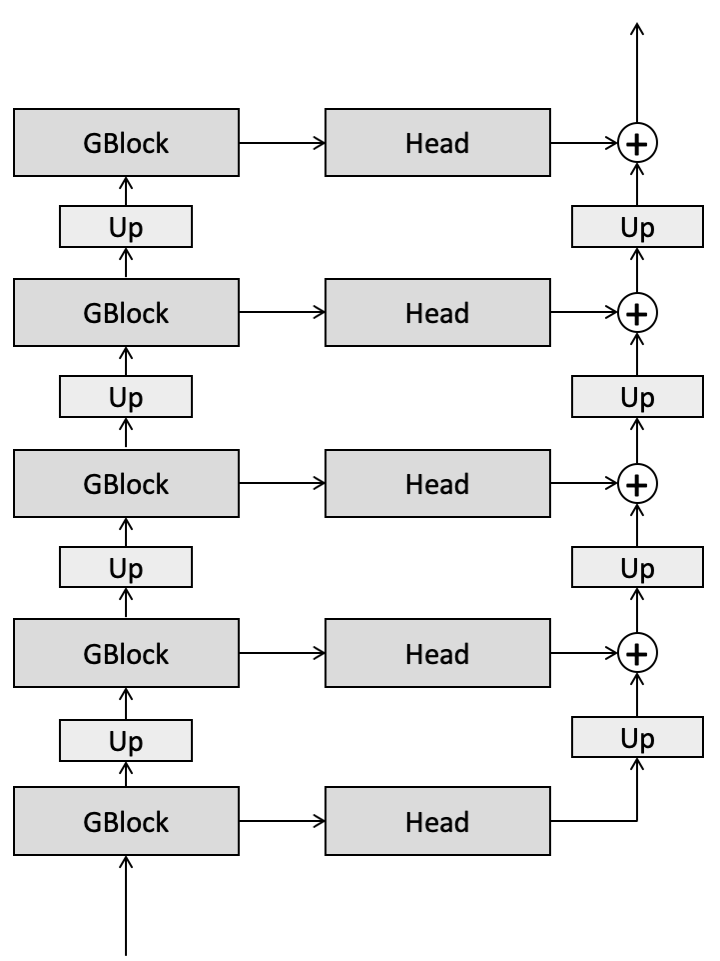}
	\caption{The  hierarchical architecture of  the proposed generator. Here, `GBlock' is a stack of convolution layers and gate recurrent unit layers with skip connections; `Head' is a convolution layer; and `Up' is a nearest-neighbor upsampling operation with scaling factor of 2.}
	\label{fig:g_arch}
\end{figure}

To alleviate the aforementioned issue, we make two changes to the architecture of the generator. First, we reduce the number of input noise vectors by a downsampling factor of $S$. That is, to generate a sequence of length $T$, we use $T'=\lceil T/S \rceil$ input noise vectors. 
Due to this change, it is necessary to upsample the output of the intermediate layers of the generator. This incurs the second change to the generator architecture---instead of a pure upsampling operation, we adopt a hierarchical architecture where the generator generates acoustic features from coarse ones to more refined ones, as shown in Figure \ref{fig:g_arch}. This kind of processing has been shown effective in several generation models \cite{Karras2019,melnet}.

As depicted in Figure \ref{fig:g_arch}, in the proposed generator we still employ the `GBlock' proposed in our prior work \cite{liu2019score}. It entails a stack of gate recurrent unit (GRU) layers \cite{cho14,chung14} and grouped convolution layers. 
Therefore, the major changes are associated with the use of the upsampling layers (`Up'), and the auxiliary convolution layers (`Head') in the skip connections.

\subsection{Cycle Regularization in the Generator}
\label{sec:reg}

Another issue of the previous model \cite{liu2019score} is that it seems to suffer from the ``mode collapse'' issue \cite{salimans16improved,Mao_2019_CVPR} and generate sounds that are not diverse enough. 
We propose using a cycle regularization mechanism in the generator to alleviate this. 

In cycleGAN \cite{cyclegan}, a cycle-consistency constraint is enforced on two generators of two different domains, so that the content generated by one generator and the content generated by the other generator have certain degree of correspondence. In our case, one domain is about the acoustic features $\mathbf{x}_t$ and the other domain the input noises $\mathbf{z}_t$. In other words, the cycle consistency is established between the input noise and the target output. We note that Ulyanov \emph{et al.} \cite{Ulyanov2018} have used this strategy as an alternative way for generation without the adversarial term. 

Specifically, besides the GAN loss $l_G$, we add the following cycle regularization term to the loss function of the generator:
\begin{equation}
l_C=|E(G(\mathbf{Z}))-\mathbf{Z}|_1 + |G(E(\mathbf{X}))-\mathbf{X}|_1 \,,
\end{equation}
where $E$, an encoder, predicts $\mathbf{Z}$ from the output of $G$. In our implementation, $E$ consists of 2D and 1D convolution layers similar to $D$, but has downsampling operations to match the upsampling operations of $G$.    
The loss function of the generator therefore becomes
\begin{equation}
l'_G = l_G + 
\lambda ~l_C \,,
\end{equation}
where $\lambda$ is a tunable hyperparameter.

If the cycle regularization term is met perfectly, it will enforce a one-one and onto mapping between the domain of noises and the domain of acoustic features. In other words, for every acoustic feature vector $\mathbf{x}$, there is a vector $\mathbf{z}$ such that $G(\mathbf{z})=\mathbf{x}$. This also means that every mode is covered, avoiding mode collapse. 
However, as the cycle term is not met perfectly in reality,  mode collapse may still happen, yet hopefully to a less degree. 

Another benefit of this regularization is that it is general-purpose: it can be augmented to any generators and discriminators without changing the architectures.

\subsection{The Discriminator}

In \cite{liu2019score}, the discriminator also contains GRUs. We have found that it is possible to use a more efficient discriminator that contains only convolution layers without compromising the performance. Specifically, we use 2D convolution layers followed by 1D convolution layers in this work, as shown in Table \ref{tab:d_arch}.

\begin{table}
\centering
\begin{tabular}{lllll}
\toprule
& {\bf Details} & {\bf Ch.} & {\bf Stride} & {\bf Dilation} \\
\toprule
{\bf 2D Block 1} & 2D Conv & 4 & (2, 1) & (1, 2)\\
& Batch norm \\
& Leaky ReLU \\
{\bf 2D Block 2} & * & 16 & (2, 1) & (1, 4) \\
{\bf 2D Block 3} & * & 64 & (2, 1) & (1, 8) \\
\midrule 
{\bf Flatten} \\
\midrule 
{\bf 1D Block 1} & 1D Conv & 512 & 1 & 1 \\
& Batch norm \\
& Leaky ReLU \\
{\bf 1D Block 2} & *        & 512 & 1 & 16 \\
{\bf 1D Block 3} & *        & 512 & 1 & 32 \\
{\bf 1D Block 4} & *        & 512 & 1 & 64 \\
{\bf 1D Block 5} & *        & 512 & 1 & 128 \\
\midrule 
{\bf Output}     & 1D Conv  & 80  & 1 & 1 \\
\bottomrule
\end{tabular}
\caption{The architecture of the discriminator. All convolution layers have kernel size 3. `Ch.' denotes the number of channels.}
\label{tab:d_arch}
\end{table}

\section{Evaluation}
\label{sec:eval}

To demonstrate that the proposed model works for a diverse set of audio, we apply it to generating singing voices, speech, and solos of piano and violin, each using a different dataset. 
Examples of the generated sounds can be found in our GitHub repo. Below, we present the
implementation details, as well as the objective and subjective studies that validate the effectiveness of the proposed model for generating singing and speech voices.





\subsection{Datasets}
\label{sec:db}
We employ the following audio datasets in this work.
\begin{itemize}
    \item \textbf{Speech}: We use the LJ Speech dataset \cite{ljspeech17}, which contains 13,100 short audio clips of a single speaker reading passages from books.
    \item \textbf{Singing}: Following our previous work \cite{liu2019score}, for the singing part we use a collection of 17.4 hours of female voices singing in Jazz. The singing voices are obtained by applying a state-of-the-art blind source separation model \cite{liu19} to the original songs which contains instrumental background music.
    \item \textbf{Piano}: We use the MAESTRO dataset  \cite{wave2midi2wave}, which contains 23 hours of virtuosic piano performance of classical music recoreded in several years. The MIDI part of the dataset has been increasingly used in symbolic-domain music generation \cite{huang2018music}. We use however the audio recordings of the data, using only those from the year of 2004. 
    \item \textbf{Violin}: We use an in-house collection of around 16.7 hours of high-fidelity violin solo recordings of classical music from various composers and violinist.
\end{itemize}

\subsection{Model Implementation Details}
\label{sec:model_detail}

For fair performance comparison with our prior model \cite{liu2019score}, we follow as closely its model settings, except for the major changes mentioned in Section \ref{sec:model}.  
We use 20-dimensional noise vectors as the input (i.e., $N=20$), and 80-dimensional mel-spectraograms as the output acoustic features (i.e., $K=80$). The parameters of the network are optimized with Adam \cite{adam} with 0.0001 learning rate. And, a mini-batch of size 5 is used and 100,000 updates are executed for each model. 
The training takes around 0.5~1 day on an NVIDIA RTX 2080Ti.

The mel-spectrograms are converted into waveforms using the MelGAN \cite{melgan} as the vocoder. 
Except for speech, we train a specific MelGAN for each of the audio types listed in Section \ref{sec:db}. For speech, we directly use the vocoder trained on LJ Speech provided in the official MelGAN GitHub repository. \footnote{\url{https://github.com/descriptinc/melgan-neurips}}

\subsection{Evaluation Metrics}
\label{sec:objective_metrics}

We employ the following objective metrics for our task.
\begin{itemize}
    \item \textbf{Vocalness} measures whether an audio clip contains human voices. 
    Following \cite{liu2019score}, we employ the JDC model 
    \cite{kum19as}\footnote{\url{https://github.com/keums/melodyExtraction_JDC}} for measuring vocalness. The JDC model regards a frame as being vocal if it has a vocal activation $\ge 0.5$ AND if the detected pitch value falls within a reasonable human pitch range (i.e., 73--988 Hz). We define the vocalness of an audio clip as the percentage of non-silent frames that are vocal.\footnote{The non-silent frames are derived by using the \texttt{librosa} function `effects.\_signal\_to\_frame\_nonsilent.'}
    \item \textbf{Diversity}. Following \cite{Mao_2019_CVPR}, we employ the following two diversity metrics  proposed by Richardson \emph{et al.} \cite{richardson18} to examine the generated mel-spectrograms: statistically-different bins (NDB) and Jensen-Shannon divergence (JSD). They measure diversity by 1) clustering the training data into several clusters, and 2) measuring how well the generated samples fit into those clusters. In other words, it uses the training data as a baseline of diversity and compares the generated samples with this baseline. We use the the official implementation of NDB and JSD.\footnote{\url{https://github.com/eitanrich/gans-n-gmms/blob/master/utils/ndb.py}}
\end{itemize}


In the subjective evaluation, we ask human listeners to rate the generated samples (on a five-point scale) by the following four metrics for singing, and by Naturalness only for speech.  
\begin{itemize}
    \item \textbf{Naturalness} measures whether an audio recording sounds like real singing voices or speeches.
    \item \textbf{Audio Quality} measures the perceptual audio quality.
    \item \textbf{Diversity} measures the  diversity of the audio contents across three different samples generated by the same model, and presented to the participant consecutively.
    \item \textbf{AI Vocalness} measures whether the generated singing voices fit the listener's own (and subjective) expectation of vocals from an AI. This metric is included with the assumption that an singing voice generating AI may have its own timbre that cannot be found in human voices.
\end{itemize}





\begin{table}[t]
\begin{center}
\begin{tabular}{lccc}
\toprule
\bf Singing         & \bf Non-Hier. & \bf  Hier. & \bf Hier. w/ cycle \\
\midrule
Naturalness     & 2.55 $\pm$ 1.07 & \bf 3.40 $\pm$ 0.92 & 2.60 $\pm$ 1.02  \\
Audio Quality   & 1.90 $\pm$ 0.70 & \bf 2.75 $\pm$ 0.94 & 2.45 $\pm$ 1.02 \\
Diversity       & 2.60 $\pm$ 0.97 & \bf 3.15 $\pm$ 0.73 & 2.95 $\pm$ 0.97 \\
AI Vocalness    & 2.75 $\pm$ 1.26 & \bf 3.40 $\pm$ 0.86 & 2.85 $\pm$ 1.15 \\ 
\midrule

\toprule
\bf{Speech} \\
\midrule
Naturalness & NA & 2.42 $\pm$ 1.02. & \bf 3.50 $\pm$ 1.06 \\

\bottomrule
\end{tabular}
\end{center}
\caption{Result of subjective evaluation on a 1-to-5 five point Likert scale; the higher the better.}
\label{tab:subjective}
\end{table}




\subsection{Subjective Evaluation Result}
\label{sec:eval_result_1}


We discuss the subjective evaluation first. We solicit non-paid responses from the Internet with an online questionnaire. 
For the singing part, a subject is asked to listen to three 10-second samples generated by a model and then rate the performance of the model based on an overall impression on the three samples. For the speech part, the subject rates (the Naturalness of) each generated 10-second sample  individually.

We compare the following three models for the singing part:
\begin{itemize}
    \item \textbf{Non-Hier.}: The old model originally proposed in \cite{liu2019score}.
    \item \textbf{Hier.}: The proposed new model using the hierarchical structure but not the cycle regularization.
    \item \textbf{Hier. w/ cycle}: The proposed model with both hierarchical structure and cycle regularization.
\end{itemize}
For the speech part, we only compare the second two models. We ask a subject to rate in total 6 generated samples, 3 from each model. The ordering of the samples are randomized.

We inform the subjects that the samples are freely generated by machine without following any text, lyrics, or pitch labels.

The responses from 20 subjects are summarized in Table \ref{tab:subjective}. We can see that the generators with a hierarchical architecture greatly outperform the non-hierarchical one in almost all the four metrics for the singing voices, demonstrating the effectiveness of the proposed hierarchical structure. Moreover, while the average rating for the old model is all under `3' (i.e., below average), this is not the case for the new model, except for Audio Quality.

Interestingly, the cycle regularization largely improves the generation quality for the speeches, but not for the singing. We conjecture that there might be two reasons for this discrepancy. First, the training data for the singing voices are the output of a source separation model, which could be noisy, while the training data for the speeches are clean speech data from a single person. The cycle regularization retain the modes but can also retain the modes of the noisy signals at the same time. Second, compared to the singing voices, speeches have a clearer target for the encoder to predict, that is, the phonemes, while the factors that are involved in singing are more complicated \cite{leeCJKL19,qian20arxiv}. 


\subsection{Objective Evaluation Result}
\label{sec:eval_result_2}





\begin{table}[t]
\begin{center}
\begin{tabular}{lccc}
\toprule
\bf Singing         & \bf Non-Hier. & \bf Hier. & \bf Hier. w/ cycle\\
\midrule
Vocalness $\uparrow$  & 0.48 $\pm$ 0.11 &  0.58 $\pm$ 0.07& \bf 0.64 $\pm$ 0.10 \\
NDB $\downarrow$      & \bf 48            & 61                & 50 \\
JSD $\downarrow$      & \bf 0.04         & 0.06             & 0.05 \\ 


\toprule
\bf{Speech} \\
\midrule
Vocalness $\uparrow$  & NA    & 0.35 $\pm$ 0.07 & \bf 0.49 $\pm$ 0.04 \\
NDB $\downarrow$      & NA    & 37                & \bf 8 \\
JSD $\downarrow$      & NA    & 0.03             & \bf 0.01 \\

\bottomrule
\end{tabular}
\end{center}
\caption{Result of objective evaluation. The Vocalness \cite{kum19as} are the higher the better, while NDB and JSD \cite{richardson18} are the opposite. }
\label{tab:objective}
\end{table}

For the objective evaluation, 100 10-second samples are generated by each model with the same random seed. We report the average scores across the 100 samples in Table \ref{tab:objective}. 

For the vocalness, we can see that the hierarchical architectures perform better than the non-hierarchical one for both singing voices and speeches. Using cycle regularization makes relatively moderate difference in the singing voices, but large improvement in speech. 
This finding seems to be consistent with the result of subjective evaluations. 

From the result of NDB and JSD, we can see that hierarchical architectures with cycle regularization improves the diversity for both singing voices and speeches. This is as expected because the cycle regularization is meant to reduce mode collapse. 

There are two interesting findings related to diversity. First, for singing, the non-hierarchical architecture actually performs the best among the three models. Our conjecture is that the hierarchical structure  enforces a consistency among local features and reduces the degree of freedom. 
This makes the diversity lower in NDB and JSD, but makes the audio more pleasant to listen to, as shown in the subjective evaluation.

Second, the singing voices with cycle regularization has better \textit{objective} diversity, while those without cycle has better \textit{subjective} diversity, comparing Tables \ref{tab:subjective} and \ref{tab:objective}. This may be related to the artifact of source separation again---the diversity caused by the artefacts might be counted by the objective metrics but not perceived as contributing to diversity subjectively.




\section{Conclusions}
\label{sec:conclusions}
In this paper, we have proposed a new  model for unconditional generation of  general audio of arbitrary length. It features a hierarchical generator to convert a sequence of random noises into slices of a mel-spectrogram, and a cycle regularizer between the noise input and the output. We have subjectively and objectively validated the effectiveness of the proposed design in different types of voices. For future work, we intend to compare the model against other existing models (e.g., \cite{melnet}), and to further improve the audio quality of the generated samples.


\bibliographystyle{IEEEtran}

\bibliography{is20}

\end{document}